\documentclass
[twocolumn,aps,prc,amsmath,amssymb,floatfix,nofootinbib]
{revtex4}
\usepackage{amssymb}
\usepackage{epsfig}
\usepackage{epstopdf}
\usepackage{graphicx}
\usepackage{bm}                      
\usepackage{mathptmx}                
\usepackage{dcolumn}                 
\usepackage{array}

\newcommand{\Msun}{\,{\rm M}_{\odot}}

\begin{document}

\title{Sound velocity in dense stellar matter with strangeness and compact stars}
\author{Chengjun~Xia$^{1}$, Zhenyu Zhu$^{2}$, Xia Zhou$^{4}$, Ang Li$^{2}$}
\affiliation{
$^1$ School of Information Science and Engineering, Zhejiang University Ningbo Institute of Technology, Ningbo 315100, China\\
$^2$ Department of Astronomy, Xiamen University, Xiamen, Fujian 361005, China; {\tt liang@xmu.edu.cn}\\
$^3$ Xinjiang Astronomical Observatory, Chinese Academy of Sciences, Urumqi, Xinjiang 830011, China
}
\date{\today}

\begin{abstract}
The phase state of dense matter in the intermediate density range ($\sim$1-10 times the nuclear saturation density) is both intriguing and unclear and could have important observable effects in the present gravitational wave era of neutron stars. As the matter density increases in compact stars, the sound velocity is expected to approach the conformal limit ($c_s/c=1/\sqrt{3}$) at high densities and should also fulfill the causality limit ($c_s/c<1$). However, its detailed behavior remains a hot topic of debate. It was suggested that the sound velocity of dense matter could be an important indicator for a deconfinement phase transition, where a particular shape might be expected for its density dependence.
In this work, we explore the general properties of the sound velocity and the adiabatic index of dense matter in hybrid stars, as well as in neutron stars and quark stars. 
Various conditions are employed for hadron-quark phase transition with varying interface tension. 
We find that the expected behavior of the sound velocity can also be achieved by the nonperturbative properties of the quark phase, in addition to a deconfinement phase transition.
And it leads to a more compact star with a similar mass. We then propose a new class of quark star equation of states, which could be tested by future high-precision radius measurements of pulsar-like objects.
\end{abstract}


\maketitle

\section{Introduction}

The equation of state (EOS) of the dense stellar matter is a mutual problem for nuclear physics and relativistic astrophysics and has been greatly promoted by the detection of gravitational waves from the GW170817 binary neutron star (NS) merger event~\cite{2017PhRvL.119p1101A,2018PhRvL.121p1101A}. 
See some recent developments reviewed e.g., in \cite{2019PrPNP.10903714B,2020GReGr..52..109C,2020JHEAp..28...19L}. 
An accurate estimation of the stars' radii ($11.9^{+1.4}_{-1.4}$ km at the $90\%$ credible level~\cite{2018PhRvL.121p1101A} was performed from the gravitational wave signal in the late inspiral stage, namely the tidal deformabilities of the stars in the binary, based on a parametrized EOS fulfilling the two-solar-mass constraint from pulsar mass measurements~\cite{2013Sci...340..448A,2010Natur.467.1081D,2016ApJ...832..167F,2018ApJS..235...37A}.
Using X-ray missions, it is also possible to simultaneously measure the masses and radii of the stars in NS low-mass X-ray binaries (LMXBs) and millisecond pulsars~\cite{2019SCPMA..6229503W}.
Recently the NICER mission has obtained the mass ($1.44^{+0.15}_{-0.14}\Msun$) and radius ($13.02_{-1.06}^{+1.24}~\rm km$) of PSR J0030+0451 to the $68.3\%$ credibility interval~\cite{2019ApJ...887L..21R,2019ApJ...887L..24M}.
Some possible implications of the measurements have also been studied combining with the gravitational-wave observations~\cite{2019ApJ...881...73W,Li2021a,Li2021b}. 
Those observations are crucial for the detailed study of the matter state at several times of nuclear saturation density $\rho_0$ (with $\rho_0=2.8\times10^{14}\rm g/cm^3$). Among them, the possibility of the existence of strange quark matter (SQM) in NSs' high-density cores is of particular interest. It could be investigated with future Advanced LIGO/Virgo detectors~\cite{2003MNRAS.338..389M,2020ApJ...904..187W, 2019MNRAS.484.4980A,2019PhRvL.122f1102B,2019PhRvL.122f1101M}.
For example, it was shown that the loss of thermodynamic convexity of EOS (or the loss of monotonicity of the sound velocity $c_s$) could have direct imprints on both the dynamics of the collapse to black hole configurations and the resulting gravitational waves~\cite{2019MNRAS.484.4980A}. 
And the sound velocity behavior is also a current pursuit in relativistic heavy-ion collisions, where some interesting findings have already been obtained~\cite{2018ARNPS..68..211N}.

\begin{table*}
\caption{\label{table:NM} The saturation properties of five nuclear matter EOS models employed which are consistent with the constraints of terrestrial experiments and nuclear theories, i.e., $K = 240 \pm 20$ MeV~\cite{2006EPJA...30...23S}, $E_{\rm sym} = 31.7 \pm 3.2$ MeV and $L = 58.7 \pm 28.1$ MeV~\cite{2013PhLB..727..276L,2017RvMP...89a5007O}. 
Also listed are the maximum gravitational mass of NSs ($M_{\rm TOV}$) and the radius of a typical $1.4 \Msun$ star.}
\setlength{\tabcolsep}{0.8pt}
\renewcommand\arraystretch{1.1}
\begin{ruledtabular}
\vspace{+0.1cm}
\begin{tabular*}{\hsize}{@{}@{\extracolsep{\fill}}ccccccccc@{}}
Model      & $n_0$        &   $E/A$    &   $K$  &  $E_{\rm sym}$   & $L$   &  $M_{\rm TOV}$ & $R_{\rm 1.4}$\\
         & (fm${}^{-3}$)  &   (MeV)    &   (MeV)  &  (MeV)  &  (MeV)  &  ($\Msun$) & (km )     \\   \hline
TW99     &  0.153       &  16.25   & 240.27 &  32.77 & 55.31  &2.09 & 12.3\\
DDME2    &  0.152       &  16.14   & 250.92 &  32.30 & 51.25  & 2.50 & 13.1\\
QMF18    &  0.16       &  16.00   & 240.00 &  31.00 & 40.00  & 2.07 & 11.9\\
BCPM     &  0.16       &  16.00   & 213.75 &  31.92 & 52.96  & 1.98 & 11.7\\
APR      &  0.16       &  16.00   & 247.30 &  33.90 & 53.80  & 2.21 & 11.4\\
\end{tabular*}
\end{ruledtabular}
\vspace{-0.5cm}
\end{table*}

Model studies on the hadron-quark EOS indicate likely a first-order quark deconfinement phase transition characterized by a decreasing behavior~\cite{2020NatPh..16..907A} 
of the adiabatic index $\Gamma=(\rho+P)(dP/d\rho)/P$. The sound velocity $c_s=\sqrt{dP/d\rho}$ should also decrease with the density but exhibits a much uncertain modification~\cite{2014ApJ...789..127K,2015PhRvL.114c1103B,2017PhRvC..95d5801M,2018MNRAS.478.1377A,2018PhRvC..98d5804T,2020NatPh..16..907A,2019PhRvD.100k4003M,2019PhRvL.122l2701M,2019AIPC.2127b0030B,2020ApJ...904..103M}
, especially at a density around $3\text{-}5\rho_0$. 
As indicated in Ref.~\cite{2018PhRvC..98d5804T}, if the two-solar-mass constraint is combined with the knowledge of hadronic matter EOS below and around nuclear saturation density, $c_s$ might first increase then decrease after reaching a maximum (maybe even up to $0.9c$ with $c$ being the velocity of light), and finally approach from below to the conformal limit $c/\sqrt{3}$, which corresponds to that of gases constitute with ultra-relativisitc massless particles. The peculiar shape  resembles the analysis in the case of the crossover EOS~\cite{2019ApJ...885...42B}. If the deconfinement phase transition is of first-order, under Maxwell construction, there is an energy density jump at transition pressure~\cite{2013PhRvD..88h3013A}, leading to $c_s=0$ and sharp peaks in the curve. This is the case if the surface tension of hadron-quark interface $\sigma$ exceeds some critical value $\sigma_c$. Under Gibbs construction, the mixed-phase consists of point-like hadron matter and quark matter~\cite{1992PhRvD..46.1274G}. 
For a moderate $\sigma$ (for example $\sim \rm 20~MeV/fm^2$ as found in Dyson-Schwinger equation approach~\cite{2016PhRvD..94i4030G})
, pasta phase with various shapes are possible~\cite{2007PhRvD..76l3015M} 
and the pressure monotonously increases with energy density.

Nevertheless, although it is known that the degree of freedom is hadron around nuclear saturation density, the QCD phase state for cold, dense matter for intermediate densities are unfortunately unknown, and a great deal of effort is undergoing in the communities of astrophysics, nuclear physics, and particle physics due to its crucial importance. One key point is still unclear: Does the matter go through a phase transition from hadron matter to quark matter at some intermediate densities, or is quark matter the absolute ground state of strongly interacting matter? (the conjecture of Bodmer-Witten-Terazawa~\cite{1971PhRvD...4.1601B,1984PhRvD..30..272W,Terazawa1979}). 
Because of the tension of a low tidal deformability ($190^{+390}_{-120}$~\cite{2018PhRvL.121p1101A}) 
and a high maximum mass ($2.14^{+0.10}_{-0.09}\Msun$ as the presently heaviest pulsar~\cite{2020NatAs...4...72C}, and $\le2.35\Msun$ based on the numerical simulation studies on NS binary mergers~\cite{2018ApJ...852L..25R,2018PhRvD..97b1501R,2019PhRvD.100b3015S}) 
for a certain EOS in the NS model, binary strange quark stars (QSs) have been proposed to be the possible scenario for the GW170817 event~\cite{2018PhRvD..97h3015Z,2018RAA....18...24L}. 
A binary QS merger for some binary configurations could eject a comparable amount of matter (to the binary NS case)~\cite{2009PhRvL.103a1101B}, to account for the electromagnetic observation in the optical/infrared/UV bands (namely kilonova). And a magnetar with QS EOS is preferred as the post-merger remnant to explain some groups of short gamma-ray burst (SGRB) observations~\cite{2016PhRvD..94h3010L,2017ApJ...844...41L}.

Therefore to understand the dense matter's phase state in intermediate densities relevant to compact stars, the present study aims to contribute a comprehensive study on the general properties of $\Gamma$, $c_s$ in the first-order quark confinement phase transition as well as in pure nuclear matter and quark matter. 
We are also interested in connecting the studies of mixed-phase and pure quark phase and establishing some quantitative results for star properties testable to observations.
The paper is organized as follows. In Sec. II, we introduce various nuclear many-body approaches employed for the hadron phase and the four effective models employed for the quark phase, including the construction of hadron-quark mixed-phase under Maxwell and Gibbs constructions as well as different hadron-quark interface tensions. Sec.  III is devoted to the discussions, before a short summary in Sec. IV.

\section{EOS models for dense matter and compact stars}

Presently we have no unified models to deal with the hadron phase and the quark phase since the matter is described under different stability conditions. The parameter space for these two states is separated. We use a pure nuclear matter model for the calculations of NSs and a pure SQM model for the calculations of QSs. The hadron-quark phase transitions are explored by combining a nuclear matter model with a quark matter model under various equilibrium conditions between two phases, then the properties of hybrid stars (HSs; namely NSs whose cores contain deconfined quarks) can be obtained.

\subsection{Nuclear matter}

For the study of nuclear matter, we choose the relativistic mean-field model (RMF) model (with the TW99~\cite{1999NuPhA.656..331T}, DDME2~\cite{2005PhRvC..71b4312L} effective interaction), the quark mean-field (QMF) model~\cite{2018ApJ...862...98Z}, the Brueckner-Hartree-Fock (BHF) approach (with the latest version BCPM~\cite{2015A&A...584A.103S}), and the variational method (with the standard Akmal-Pandharipande-Ravenhall (APR) formalism~\cite{1998PhRvC..58.1804A}). They are among various microscopic calculations or most-advanced and widely-employed phenomenological models, without much dependence on the model parameters for the results being as general as possible.

At densities below and around the nuclear saturation density $n_0\approx 0.15\text{-}0.16\ \mathrm{fm}^{-3}$, the EOS of nuclear matter is well constrained with terrestrial experiments and nuclear theories, which gives the energy per baryon $E/A\approx 16$ MeV, the incompressibility $K = 240 \pm 20$ MeV~\cite{2006EPJA...30...23S}, the symmetry energy $E_{\rm sym} = 31.7 \pm 3.2$ MeV and its slope $L = 58.7 \pm 28.1$ MeV~\cite{2013PhLB..727..276L,2017RvMP...89a5007O}.\footnote{Note that the recent PREX-II results~\cite{2021arXiv210103193R} on the neutron skin thickness of $^{208}\mathrm{Pb}$ may indicate a $L$ value about twice the previous one.}
The saturation properties of nuclear matter for the the employed five EOS models are collected in Table~\ref{table:NM}, together with maximum mass of a pure NS and the radius of a typical $1.4 \Msun$ star. We mention that five NS EOS models all fulfill the available robust mass/radius measurements from the gravitational wave signal and electromagnetic signals~\cite{2018PhRvL.121p1101A,2013Sci...340..448A,2010Natur.467.1081D,2016ApJ...832..167F,2018ApJS..235...37A,2019ApJ...887L..21R,2019ApJ...887L..24M}.

\subsection{Quark matter}
\label{sec:qm}

The SQM is composed of up ($u$), down ($d$) and strange ($s$) quarks with the charge neutrality maintained by the inclusion of electrons (hereafter muons as well if present),
\begin{equation}
\frac{2}{3}n_u-\frac{1}{3}n_d-\frac{1}{3}n_s-n_e=0 \:,  \label{eq:Chargeneut}
\end{equation}
The baryon number conservation,
\begin{equation}
\frac{1}{3}\left(n_u + n_d + n_s\right) =n_{\rm b} \:, \label{eq:Baryonconserv}
\end{equation}
is also satisfied with $n_{\rm b}$ being the baryon number density.
Due to the weak interactions between quarks and leptons,
\begin{eqnarray}
&& d \rightarrow u + e + \tilde{\nu}_e \:,~~ u + e \rightarrow d + \nu_e \:;\nonumber \\
&& s \rightarrow u + e + \tilde{\nu}_e \:,~~ u + e \rightarrow s + \nu_e \:;\nonumber \\
&& s + u \leftrightarrow d + u \:, \nonumber
\end{eqnarray}
and the $\beta$-stable conditions $\mu_s = \mu_d = \mu_u + \mu_e $ should be fulfilled.
The energy density and pressure include both contributions from quarks and leptons, and those of leptons can be easily calculated by the model of ideal Fermi gas. In this section, we mainly introduce the necessary formalism for quarks.

In the density regime achieved inside compact stars, it is not applicable for the dense matter properties to be calculated directly from the first principle lattice quantum chromodynamics (QCD) or from perturbative QCD. We make use of various phenomenological descriptions of the system, and our studies for SQM and quark stars are based on four effective models. The four quark matter models may include all possible QS models in the market with a high maximum mass (above $\sim 2 \Msun$) and cover approximately the full preferred radius range ($\sim10\text{-}14~\rm km$) of a typical $1.4 \Msun$ mass star. In the following, we introduce the four quark matter models, namely the MIT bag model, the perturbation model, the equivparticle model, and the quasiparticle model.

\subsection{\label{sec:the_bag} MIT bag model ($B_{\rm eff},a_4$)}

The most popular approach to obtain the properties of SQM is the MIT bag model~\cite{1986ApJ...310..261A, 1986A&A...160..121H}, with the usual correction $\sim\alpha_\mathrm{s}$ from perturbative QCD. The $O(\alpha^2_\mathrm{s})$ pressure was evaluated and approximated~\cite{2001PhRvD..63l1702F} 
in a similar simple form with the
original bag model, and was used to study hybrid stars and quark stars~\cite{2017ApJ...844...41L,2005ApJ...629..969A,2011ApJ...740L..14W,2016MNRAS.457.3101B,Li2021b}. 
At given chemical potential $\mu_i~(i=u,d,s)$, the pressure $P$, particle number density $n_i$, and energy density $\rho$ are determined by
\begin{eqnarray}
P &=& -\Omega_0 - \frac{3\mu^4}{4\pi^{2}}(1-a_4) - B_{\rm eff} \:, \label{eq:pressure_bag}\\
n_i &=&  \frac{g_i}{6\pi^2} \left(\mu_i^2-m_{i}^2\right)^3 - \frac{\mu^3}{\pi^2}(1-a_4) \:, \label{eq:ni_bag}\\
\rho &=& \sum_i \mu_i n_i - P \:, \label{eq:E_bag}
\end{eqnarray}
where the average chemical potential is $\mu= \sum_i\mu_i/3$ and $g_i$ is the degeneracy factor for particle type $i$ ($g_u=g_d=g_s=6$). The $a_4$ parameter is commonly taken to be $2\alpha_\mathrm{s}/\pi$ to one loop order~\cite{1986ApJ...310..261A, 1986A&A...160..121H} with $\alpha_\mathrm{s}$ being the strong coupling constant. 
Here both $B_{\rm eff}$ and $a_4$ are effective parameters including non-perturbative effects of the strong interactions. $\Omega_0$ takes the form of a thermodynamic potential density with non-interacting particles ($m_{u} = m_{d} = 0, m_{s} = 100$ MeV are usually used for simplicity), i.e.,
\begin{widetext}
\begin{equation}
\Omega_0 = -\sum_i\frac{g_i}{24\pi^2} \left[ \mu_i(\mu_i^2-\frac{5}{2}m_i^2)\sqrt{{\mu_i}^2-m_i^2}
           +\frac{3}{2} m_i^4\ln\frac{\mu_i+\sqrt{{\mu_i}^2-m_i^2}}{m_i}  \right] \:. \label{eq:Omega0}
\end{equation}
\end{widetext}

\subsection{\label{sec:the_pQCD} Perturbation model ($C_1, B_0, \Delta\mu$)}

As mentioned above, the property of quark matter at intermediate densities is not attainable directly by solving QCD.
The perturbative QCD can only be applicable at ultra-high densities above $\sim40\rho_0$~\cite{2014ApJ...789..127K,2014ApJ...781L..25F}.
We make use of the perturbative calculations in the present perturbation model, and introduce additionally the non-perturbative corrections through model parameters. 

We employ the pQCD thermodynamic potential density to the order of $\alpha_\mathrm{s}$~\cite{2005PhRvD..71j5014F}, i.e.,
\begin{equation}
\Omega^\mathrm{pt} = \Omega_0 + \Omega_1 \alpha_\mathrm{s} \:, \label{eq:omegapt_pQCD}
\end{equation}
with
\begin{widetext}
\begin{equation}
\Omega_1 =\sum_{i=u,d,s} \frac{g_i m_i^4}{12\pi^3}
               \left\{ \left[ 6 \ln\left(\frac{\bar{\Lambda}}{m_i}\right) + 4 \right]\left[u_i v_i - \ln(u_i+v_i)\right]
               + 3\left[u_i v_i - \ln(u_i+v_i)\right]^2 - 2 v_i^4 \right\} \:,
\label{eq:omega1}
\end{equation}
\end{widetext}
where $u_i \equiv \mu_i/m_i$ and $v_i \equiv \sqrt{u_i^2-1}$. Note that the thermodynamic potential density to the zeroth order $\Omega_0$ is the same as Eq.~(\ref{eq:Omega0}).
The coupling constant $\alpha_\mathrm{s}$ and quark masses $m_i$ are running with the energy scale and can be determined by~\cite{2005PhRvD..71j5014F}:
\begin{eqnarray}
\alpha_\mathrm{s}(\bar{\Lambda})
  &=& \frac{1}{\beta_0 L}   \left(1- \frac{\beta_1\ln{L}}{\beta_0^2 L}\right) \:,
\label{eq:alpha} \\
m_i(\bar{\Lambda})
  &=& \hat{m}_i \alpha_\mathrm{s}^{\frac{\gamma_0}{\beta_0}}
      \left[ 1 + \left(\frac{\gamma_1}{\beta_0}-\frac{\beta_1\gamma_0}{\beta_0^2}\right) \alpha_\mathrm{s} \right] \:.
\label{eq:mi}
\end{eqnarray}
Here $L\equiv \ln\left( \frac{\bar{\Lambda}^2}{\Lambda_{\overline{\mathrm{MS}}}^2}\right)$ with $\Lambda$ being the renormalization scale. 
We take the $\overline{\mathrm{MS}}$ renormalization point $\Lambda_{\overline{\mathrm{MS}}} = 376.9$ MeV based on the latest results for strong coupling constant~\cite{2014ChPhC..38i0001O}. 
Following Eq.~(\ref{eq:mi}) the invariant quark masses are $\hat{m}_u= 3.8$ MeV, $\hat{m}_d = 8$ MeV, and $\hat{m}_s = 158$ MeV.
The parameters for the $\beta$-function and $\gamma$-function are $\beta_0=\frac{1}{4\pi}(11-\frac{2}{3}N_\mathrm{f})$, $\beta_1=\frac{1}{16 \pi^2} (102-\frac {38}{3} N_\mathrm{f})$,
$\gamma_0=1/\pi$, and $\gamma_1=\frac{1}{16\pi^2} (\frac{202}{3} - \frac{20}{9}N_\mathrm{f})$~\cite{1997PhLB..405..327V} 
(The formulas is for arbitrary $N_\mathrm{f}$ and in this study $N_\mathrm{f}=3$). It is not clear how the renormalization scale evolves with the chemical potentials of quarks, and we adopt $\bar{\Lambda} = \frac{C_1}{3} \sum_i\mu_i$, with $C_1=1 \sim 4$~\cite{2014ApJ...781L..25F}.

To account for the energy difference between the physical vacuum and perturbative vacuum, we introduce the bag mechanism with a dynamically-scaled bag parameter~\cite{2002PhLB..526...19B,2004PhRvD..70d3010M}. 
The total thermodynamic potential density for SQM can be written as~\cite{2019PhRvD..99j3017X}
\begin{eqnarray}
\Omega  &=& \Omega^\mathrm{pt} + B \nonumber\\
& \equiv& \Omega^\mathrm{pt} + B_\mathrm{QCD} \nonumber\\
&& + (B_0 - B_\mathrm{QCD})\exp{\left[-\left( \frac{\sum_i\mu_i-930}{\Delta\mu}\right)^4\right]} \:. \label{eq:omega_pQCD}
\end{eqnarray}
Following~\cite{1975PhRvD..12.2060D}, we take $B_0=40,~50~\rm MeV/fm^3$ for the calculations.
$\Delta\mu=\infty$ indicates no medimum effect for the bag parameter. 
If $\alpha_\mathrm{s}$ and $m_{u,d,s}$ are running with the energy scale as reported by Particle Data Group~\cite{2014ChPhC..38i0001O}, the maximum mass of QSs does not reach $\sim2 \Msun$. 
In such cases, the dynamical rescaling of the bag constant with finite $\Delta \mu$ is essential, which basically originates from the nonperturbative effects such as chiral symmetry breaking and color superconductivity~\cite{2005PhR...407..205B,2008RvMP...80.1455A,2018RPPh...81e6902B}. 

At given chemical potentials $\mu_i$, the pressure $P$, particle number density $n_i$, and energy density $\rho$ are determined by
\begin{eqnarray}
P &=& -\Omega \:, \label{eq:pressure_pQCD}\\
n_i &=& \frac{g_i}{6\pi^2} \left(\mu_i^2-m_i^2\right)^3
          -\frac{\partial \Omega_1}{\partial \mu_i}
           \alpha_\mathrm{s} -\frac{\partial B}{\partial \mu_i} \nonumber\\
    &&{}  -\frac{C_1}{3}\sum_i \left( \frac{\partial \Omega_0}{\partial m_i}
                          +\frac{\partial \Omega_1}{\partial m_i}\alpha_\mathrm{s}\right) \frac{\mbox{d} m_i}{\mbox{d} \bar{\Lambda}}\nonumber\\
    &&{}  -\frac{C_1}{3} \frac{\partial \Omega_1}{\partial\bar{\Lambda}} \alpha_\mathrm{s}
                          -\frac{C_1}{3} \Omega_1 \frac{\mbox{d} \alpha_\mathrm{s}}{\mbox{d} \bar{\Lambda}} \:, \label{eq:ni_pQCD}\\
\rho &=& \Omega + \sum_i \mu_i n_i \:. \label{eq:E_pQCD}
\end{eqnarray}

\begin{figure*}
\begin{center}
\includegraphics[width=0.49\textwidth]{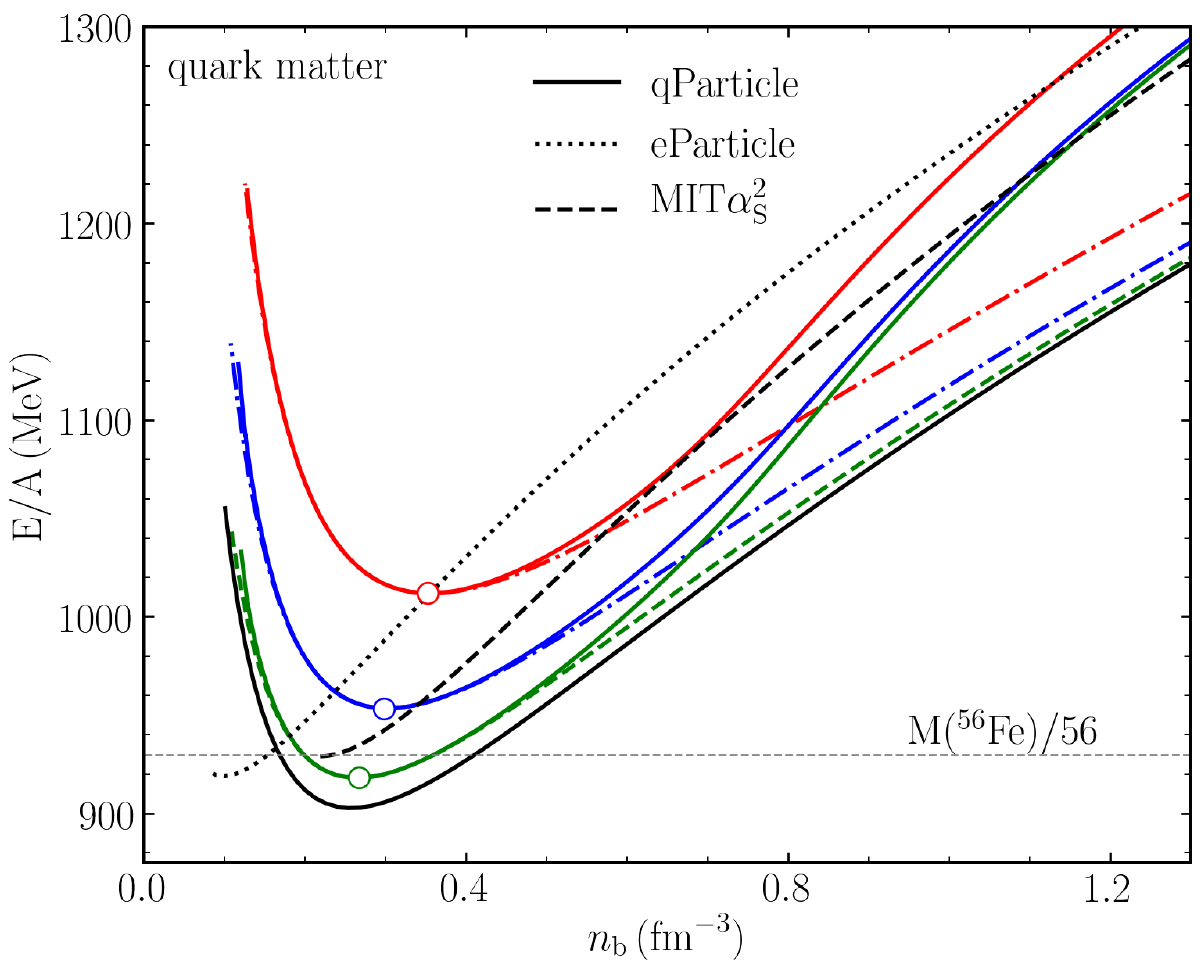}
\includegraphics[width=0.49\textwidth]{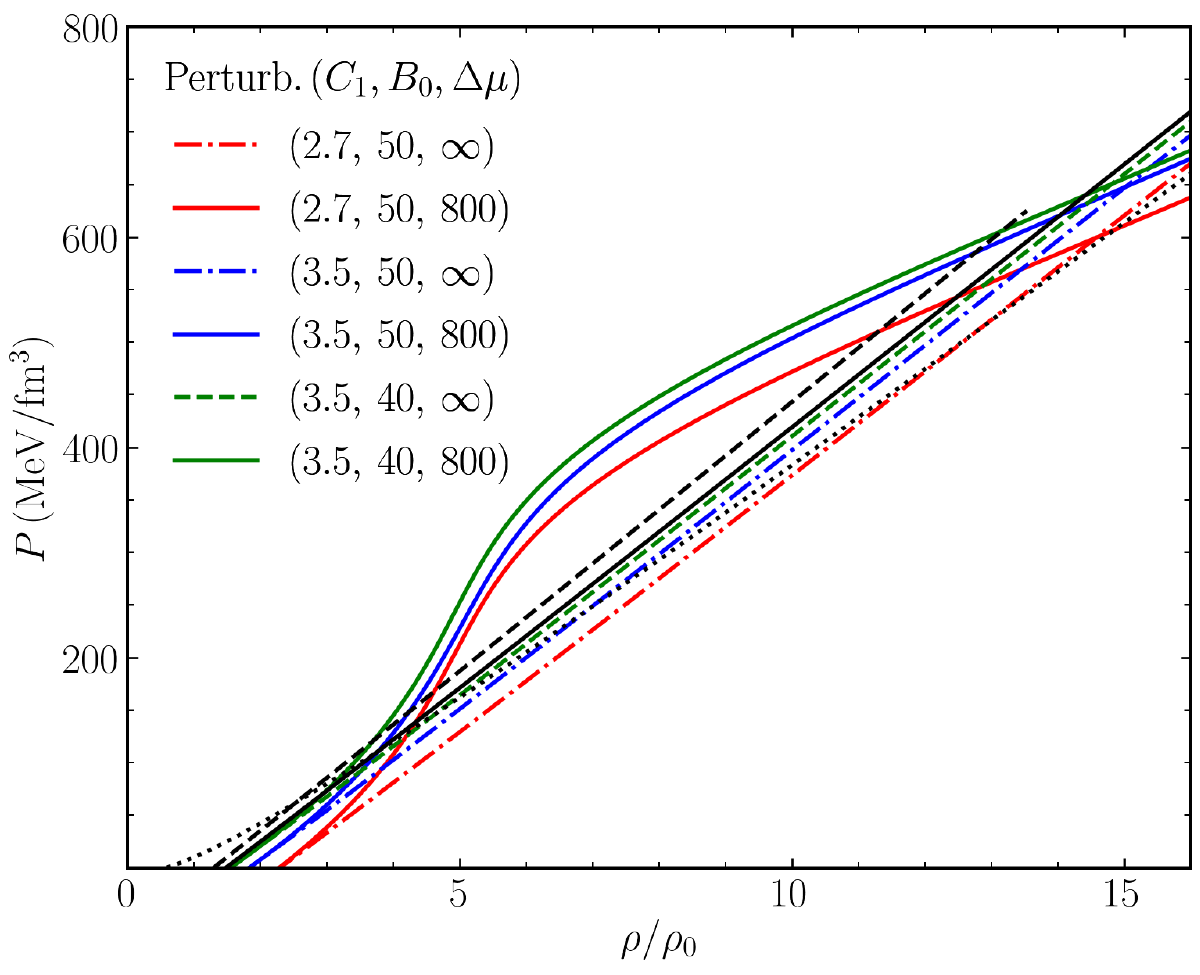}
\caption{Energy per baryon $E/A$ as a function of the baryon number density $n_{\rm b}$. The calculations are done with various effective SQM model: the quasiparticle model (black solid curves) with $C_1=3.5$, $B=50~{\rm MeV/fm^3}$, the equivparticle model (black dotted curves) with $C=0.7,~\sqrt{D}=129~\rm MeV$, the MIT$\alpha_\mathrm{s}^2$ bag model (black dashed curves) with $B_{\rm eff}^{1/4}=138~{\rm MeV}$ (namely $B_{\rm eff}\sim47.2 ~{\rm MeV/fm^3}),~a_4=0.61$, and the pertrubation model (colorfull curves) with six sets of parameter ($C_1,~B_0,~\Delta \mu$). Three dots in the left panel represent the mimimum energy points, respectively. The horizontal line corresponds to $E/A=930~\rm MeV$, which is the energy per baryon of the stablest atomic nuclei $^{56}\rm Fe$.}\label{fig:epape}
\end{center}
\vspace{-1cm}
\end{figure*}

\subsection{\label{sec:the_equiv}Equivparticle model ($C, \sqrt{D}$)}

Besides the bag mechanism, quark confinement can be achieved via density dependence of the mass, as done in the equivparticle model~\cite{2000PhRvC..62b5801P,2014PhRvD..89j5027X}. Take into account both the linear confinement and leading-order perturbative interactions, the quark mass scaling is given by
\begin{equation}
  m_i(n_{\mathrm b})= m_{i0}+Dn_{\mathrm b}^{-1/3}+Cn_{\mathrm b}^{1/3} \:,  \label{Eq:mnbC}
\end{equation}
where $m_{i0}$ is the current mass ($m_{u0} \sim2.3$ MeV, $m_{d0} \sim4.8$ MeV, $m_{s0} \sim95$ MeV)~\cite{2014ChPhC..38i0001O} 
and $n_{\mathrm b}= (n_u+n_d+n_s)/3$ is the baryon number density. The parameters $D$ and $C$ characterize the strengths of confinement and leading-order perturbative interactions, which have been estimated with 
$140 \lesssim \sqrt{D}\lesssim 270$ MeV~\cite{2005PhRvC..72a5204W} 
and $C\lesssim 1.2$~\cite{2014PhRvD..89j5027X}.

At given particle number densities $n_i$, the energy density $\rho$, chemical potential $\mu_i$ and pressure $P$ are given by
\begin{eqnarray}
\rho &=& \sum_{i} \frac{g_i}{16\pi^{2}} \left[\nu_i(2 \nu_i^2+m_i^2)\sqrt{\nu_i^2+1}-m_i^4 \mathrm{arcsh}\left(\frac{\nu_i}{m_i}\right) \right] \:, \nonumber \\
\label{eq:E_eqv} \\
\mu_i &=& \sqrt{\nu_i^2+m_i^2} + \frac{1}{9}\left(\frac{C}{n_{\mathrm b}^{2/3}}-\frac{D}{n_{\mathrm b}^{4/3}}\right) \sum_in^s_i, \label{eq:chem_eqv}\\
P &=& \sum_i \mu_i n_i  - \rho \:, \label{eq:pressure_eqv}
\end{eqnarray}
with the scalar and vector densities
\begin{eqnarray}
&&n^s_i=\langle \bar{\Psi}_i \Psi_i\rangle = \frac{g_i m_i}{4\pi^{2}} \left[ \nu_i \sqrt{\nu_i^2+1} - m_i^2 \mathrm{arcsh}\left(\frac{\nu_i}{m_i}\right) \right] \:, \nonumber \\
\\
&&n_i=\langle \bar{\Psi}_i \gamma^0 \Psi_i\rangle =  \frac{g_i\nu_i^3}{6\pi^2} \:.\label{eq:ni_eqv}
\end{eqnarray}
Here $\nu_i$ is the Fermi momentum for particle type $i$.

\subsection{\label{sec:the_quasi}Quasiparticle model ($C_1, B_0$)}

Similar to the equivparticle model, in quasiparticle model the strong interactions is mimicked by effective masses. At zero temperature, by resuming one-loop self energy diagrams in the hard dense loop approximation, the effective mass formula for quarks at finite chemical potentials can be obtained as~\cite{1989NuPhA.498..423P,1997JPhG...23.2051S,1997NuPhA.616..659S}
\begin{equation}
m_i=\frac{m_{i0}}{2}+\sqrt{\frac{m_{i0}^2}{4}+\frac{2 \alpha_\mathrm{s}}{3 \pi} \mu_i^2} \:. \label{eq:m_quasi}
\end{equation}
Here $m_{i0}$ is the current mass of quark flavor $i$~\cite{2014ChPhC..38i0001O} and $\alpha_\mathrm{s}$ the running strong coupling constant given by Eq.~(\ref{eq:alpha}).

At given chemical potentials $\mu_i$, the pressure $P$, particle number density $n_i$, and energy density $\rho$ are then determined by
\begin{eqnarray}
P &=& -\Omega = -\Omega_0 - B_0 \:, \label{eq:pressure_quasi}\\
n_i &=&  \frac{g_i}{6\pi^2} \left(\mu_i^2-m_i^2\right)^3 - \sum_{j=u,d,s} \frac{\partial \Omega_0}{\partial m_j}\frac{\mbox{d} m_j}{\mbox{d} \mu_i} \:, \label{eq:ni_quasi}\\
\rho &=& \Omega_0 +  B_0 + \sum_i \mu_i n_i \:. \label{eq:E_quasi}
\end{eqnarray}
Again the bag constant $B_0$ represents the vacuum pressure. Based on Eq.~(\ref{eq:Omega0}), the derivative of $\Omega_0$ with respect to the effective quark mass $m_i$ is calculated as 
\begin{equation}
\frac{\partial\Omega_0}{\partial m_i} = \frac{g_im_i}{4\pi^2} \left[ \mu_i \sqrt{\mu_i^2-m_i^2} -m_i^2\ln\frac{\mu_i+\sqrt{\mu_i^2-m_i^2}}{m_i} \right] \:. \label{eq:dOmega0dm}
\end{equation}

In the left (right) panel of Fig.~\ref{fig:epape} we present the energy per baryon (pressure) obtained with various effective models for representative parameters: the quasiparticle model (labeled as qParticle), the equivparticle model (labelled as eParticle), the MIT$\alpha_\mathrm{s}^2$ bag model, and the perturbation model (labelled as Pertrub.). We notice the opposite effect of $C_1$ and $B_0$ parameter on the EOS in the perturbation model, namely larger bag constant $B_0$ usually results in softening, while larger dimensionless parameter $C_1$ (namely lager renormalization scale) results in stiffening. The dynamic scaling of the $B$ parameter with a finite $\Delta \mu$ brings further repulsion and increases the energy (pressure) evidently from around $0.5~\rm fm^{-3}$ ($\sim4\rho_0$) in the left (right) panel.

To estimate whether the SQM is absolute stable strong-interaction system, we have to require at $P=0,~E/A \leq M(^{56}\rm Fe)/56=930~MeV$. The condition is fulfilled under four cases of our calculations, $\mathrm{qParticle}~(C_1,B_0)=(3.5,~50)$, $\mathrm{eParticle}~(C,\sqrt{D})=(0.7,~129)$, $\mathrm{MIT\alpha_\mathrm{s}^2}~(B_{\rm eff},a_4)=(138,~0.61)$ and $\mathrm{Pertrub.}~(C_1,B_0, \Delta\mu)=(3.5,~40,~800/\infty$). Those are the cases where a strange QS is possible and the prediction for the star properties will be presented later in this section. Since the zero-pressure density closely related to the QS EOS’s stiffness (ever can be regarded as the characteristic of the stiffness in many previous studies~\cite{2016MNRAS.457.3101B,2017ApJ...844...41L}), we mention that the surface density is the lowest in the $\rm eParticle~(0.7,~129)$ EOS, around $0.1~\rm fm^{-3}$. Its stiffness will be manifested later in the results of the star properties (Sec. \ref{sec:mr}).

\section{\label{sec:mix}Hadron-quark phase transition of first order}

To construct the hadron-quark mixed phase at two extreme scenarios with $\sigma\rightarrow 0$ (the Gibbs construction) and $\sigma>\sigma_\mathrm{c}$ (the Maxwell construction), we define the fraction of quark matter as $\chi\equiv V_q/V$, where $V_q$ is the volume occupied by quarks and $V$ the total volume, i.e., $\chi=0$ represents the pure nuclear matter and $\chi=1$ the quark matter. The total baryon number density is
\begin{equation}
n_\mathrm{b} = (1-\chi) (n_p + n_n) +\chi \left(n_u + n_d + n_s\right)/3 \:, \label{eq:nt}
\end{equation}
The total energy density is
\begin{equation}
 \rho = (1-\chi) \rho_N  +\chi \rho_q  +\rho_e \:, \label{eq:Et} \\
\end{equation}
where $\rho_N$, $\rho_q$, and $\rho_e$ are the energy densities for nuclear matter, quark matter and electrons.

The constituent particle chemical potentials in the two sector is linked as follows, $\mu_n = \mu_u + 2\mu_d,~\mu_p = 2\mu_u + \mu_d,~\mu_e=\mu_n-\mu_p=\mu_d-\mu_u$. Two independent chemical potentials $(\mu_n, \mu_p)$ or $(\mu_u, \mu_d)$ can be determined by solving the charge neutrality equation and the pressure balance equation for a given total baryon number or a given quark fraction~\cite{2008IJMPE..17.1635L,2009ChPhC..33...61L,2015PhRvC..91c5803L,2008PhRvC..77f5807P}. 
The EOS of mixed phase can be then calculated. We mention that the local charge neutrality condition,
\begin{equation}
n_p-n_e=0 \:,~\frac{2}{3}n_u-\frac{1}{3}n_d-\frac{1}{3}n_s-n_e=0\:,
\end{equation}
is fulfilled within the Maxwell phase transition construction, and the global charge neutrality condition is satisfied within the Gibbs phase transition construction,
\begin{equation}
  0 = (1-\chi)n_p +\chi \left(\frac{2}{3}n_u-\frac{1}{3}n_d-\frac{1}{3}n_s\right)-n_e \:. \label{eq:Qt}
\end{equation}

For the cases with moderate surface tension ($0<\sigma<\sigma_\mathrm{c}$), to construct the geometrical structures of mixed-phase, we employ a Wigner-Seitz approximation and assume spherical symmetry, i.e., only the droplet and bubble phases are considered. The internal structure of the Wigner-Seitz cell is determined by minimizing
the energy at a given number density. More formulas can be found in our previous study~\cite{2019PhRvD..99j3017X}.

\begin{figure*}
\begin{center}
\includegraphics[width=0.95\textwidth]{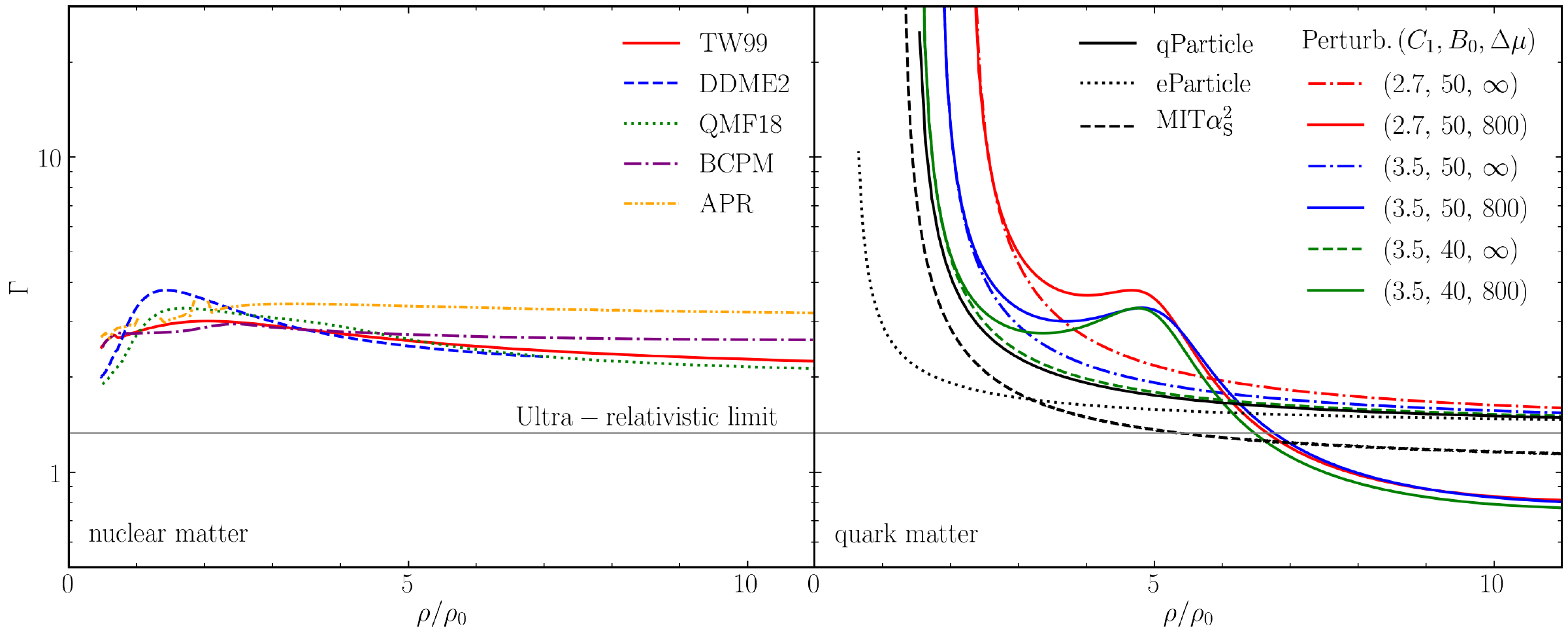}
\caption{(Left) Adiabatic index $\Gamma$ of nuclear matter (left) and of SQM (right), as functions of the energy density $\rho$ (divided by the saturation density $\rho_0$). The results of nuclear matter are obtained with five various EOS models, namely TW99, DDME2, QMF18, BCPM, APR. The calculations of SQM are done with various effective models: quasiParticle, equivparticle, MIT$\alpha_\mathrm{s}^2$, perturbation models using six sets of parameter ($C_1,~B_0,~\Delta \mu$). The horizontal line represents for the ultra-relativistic limit.}\label{fig:gamma}
\end{center}
\vspace{-0.5cm}
\end{figure*}

\begin{figure*}
\begin{center}
\includegraphics[width=0.95\textwidth]{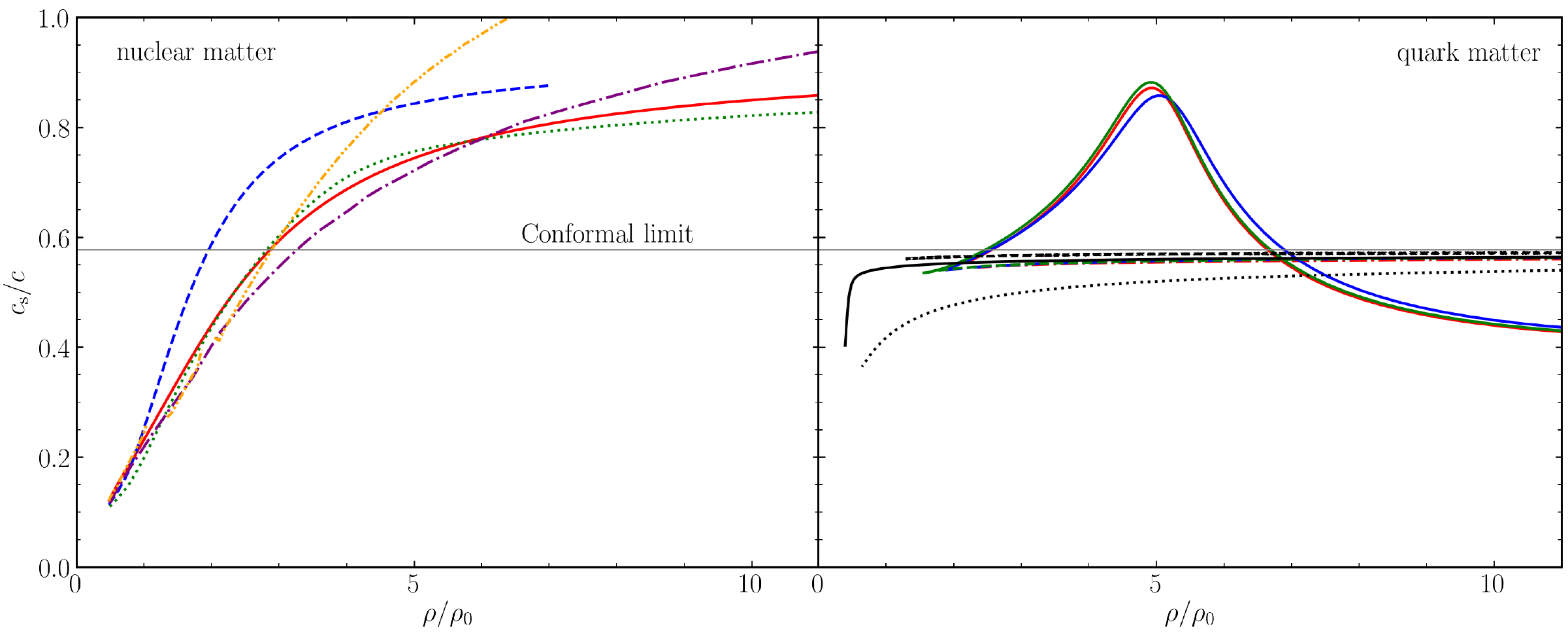}
\caption{Same with~\ref{fig:gamma}, but for the sound speed $c_s$. The horizontal line represents for the conformal limit.}\label{fig:vs}
\end{center}
\vspace{-0.8cm}
\end{figure*}

\section{Discussions}

\begin{figure*}
\begin{center}
\includegraphics[width=0.95\textwidth]{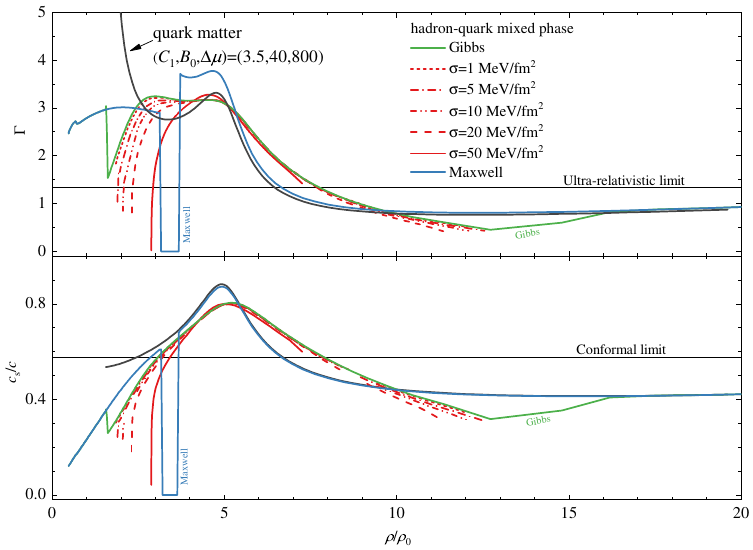}
\caption{$\Gamma$ (Upper) and $c_s$ (lower) for hybrid matter under various constructions between the two phases: Maxwell construction (shadow regions show the finite density jump in this case), Gibbs construction, and some choices of hadron-quark interface tension ($\sigma=~1,~5,~10,~20,~50~\rm MeV/fm^2$). For the calculations, the nuclear matter EOS employs the RMF model with the TW99 effective interaction, and the SQM EOS employs the perturbation model with the parameters of $C_1=2.7,~B_0=50~\rm MeV/fm^3,~\Delta \mu=800~\rm MeV$. The $c_s$ result of pure SQM case of Perturb.(3.5, 40, 800) are also shown in the lower panel for comparison. The horizontal lines in the upper/lower panel shows the ultra-relativisitc limit/conformal limit for $\Gamma$/$c_s$.}\label{fig:hybrid}
\end{center}
\vspace{-0.8cm}
\end{figure*}

\subsection{Adiabatic index $\Gamma$ and sound velocity $c_s$: nuclear matter vs. quark matter}
The results of $\Gamma$, $c_s$ for NS matter (or betastable nuclear matter) are presented in the left panel of Fig.~\ref{fig:gamma} and Fig.~\ref{fig:vs}, respectively.
The corresponding results of quark matter are shown in the right panels.

In Fig.~\ref{fig:gamma} for $\Gamma$, we see that the value mostly lies between $\sim 2\text{-}3$ for nuclear matter and commonly greater than those of SQM in the intermediate density range. The adiabatic index of SQM matter shows a sharp decrease with density. In the cases of quasiparticle model, equivparticle model, and perturbation model (with a fixed bag parameter) they also approach close to the ultra-relativistic limit of $4/3$ at high densities. The lower curves at a high-density range in the MIT$\alpha_\mathrm{s}^2$ model and perturbation model (with in-medium bags) indicate the quark interactions are weaker in these cases. In particular, we notice the stiffening of the adiabatic index in the perturbation model from the repulsive contribution brought by the dynamic scaling of the bag parameters. 

In Fig.~\ref{fig:vs} of the velocity of sound $c_s$ , we see that $c_s$ increases monotonously from small values with the density using only a nuclear matter EOS, and there is possible a violation of the causality at some high densities, for example, $\sim 6.45\rho_0$ in the APR case. The model can certainly not be applied for the study of dense matter beyond this density. We mention that the NS central density with a maximum mass of $\sim2.2\Msun$ for APR is high up to $\sim9.75\rho_0$, which is beyond the causality violation density.
For the SQM EOSs (except the perturbation model with in-medium bags), $c_s$ also increases monotonously from small values but approaches quickly (around $\rho_0$) to the conformal limit of $c/\sqrt{3}$ from below. However, for the perturbation model with in-medium bags at $\Delta \mu=800~\rm MeV$, $c_s$ increases and then decreases, resulting in a peak in the curve located $\sim5\rho_0$. This may be what expected in \cite{2015PhRvL.114c1103B} for NSs, from the analysis based on the two-solar-mass constraint and the empirical evidence below and around nuclear saturation density.
The peak can be as high as $0.9c$, similar to the result in \cite{2018PhRvC..98d5804T}. In \cite{2014ApJ...789..127K,2018MNRAS.478.1377A}, a relatively lower peak value ($\sim0.63c$) is found. The hadron-quark phase transition can achieve similar $c_s$ shapes as we will immediately show.

\begin{figure*}
\begin{center}
\includegraphics[width=0.95\textwidth]{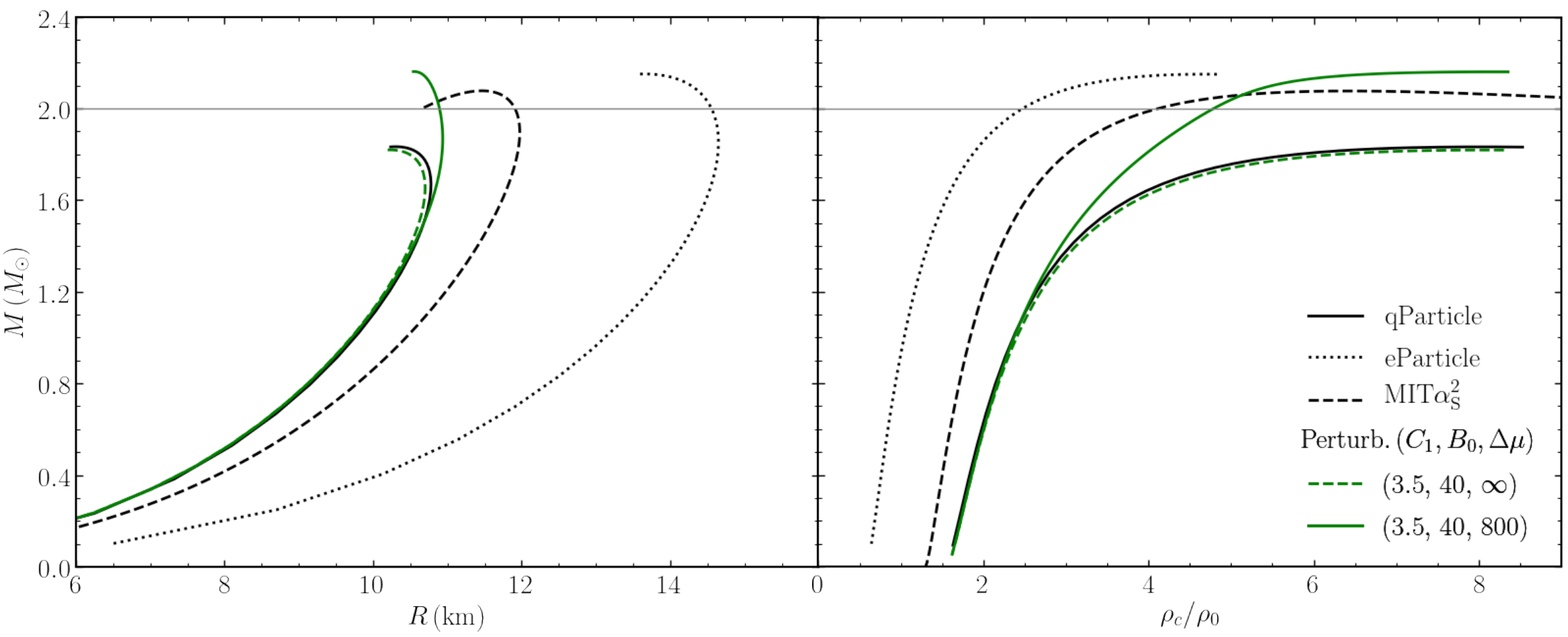}
\caption{\label{fig:mr} QSs' mass as a function of the radius (left panel) and as a function of the central density (right panel). The calculations are done 
for all four effective quark matter models with typical parameter sets, i.e., the quasiparticle model with $C_1=3.5$, $B=50~{\rm MeV/fm^3}$, the equivparticle model with $C=0.7,~\sqrt{D}=129~\rm MeV$, the MIT$\alpha_\mathrm{s}^2$ bag model with $B_{\rm eff}^{1/4}=138~{\rm MeV}$ (namely $B_{\rm eff}\sim47.2 ~{\rm MeV/fm^3}),~a_4=0.61$, and the pertrubation model with the parameters of $C_1=3.5,~B_0=40~\rm MeV/fm^3,~\Delta \mu=800~\rm MeV$ or $\infty$.
The horizontal lines indicate the two-solar-mass limit.}
\end{center}
\vspace{-0.8cm}
\end{figure*}

\subsection{Pure quark matter masquerades as mixed phase in $\Gamma,~c_s$}
\label{sec:qmvs.mixed}

We show in Fig.~\ref{fig:hybrid} the $\Gamma,~c_s$ results of the HS matter under hadron-quark phase transition.
The calculations are done using the perturbation model for quark matter, combining with two nuclear matter EOS models (soft TW99 and stiff DDME2) for the study of stiffness effects.
The results of three other quark matter models should be similar to those of the perturbation model in the cases without the dynamical scaling of the bag parameter $\Delta\mu=\infty$.
The $c_s$ result of pure SQM case of Perturb.(3.5, 40, 800) are also shown in the lower panel for comparison. The calculations are done under various constructions between the two phases: Maxwell construction (with a considerable interface tension and a finite density jump), Gibbs construction (with a zero interface tension), and some proper choices of hadron-quark interface tension (in the range of $1\text{-}50~\rm MeV/fm^2$). Varying the surface tension basically indicates that the properties of quark-hadron mixed-phase interpolate in between the two extremes, i.e., the Gibbs construction scenarios with point-like hadronic matter and quark matter and the Maxwell construction scenarios with bulk separation of the two phases. If we increase the hadron-quark interface tension, the obtained results evolve from the Gibbs case into the Maxwell case, where the density range of the mixed-phase also shrinks.

Let's first focus on the Gibbs case with no interface tension. At the quark threshold density, the adiabatic index $\Gamma$ sharply decreases by almost a factor of two due to the strongly softening of the EOS by an extra degree of freedom. Then as the density increases, $\Gamma$ grows because the pressure increases. After reaching a maximum of $\sim3.2$, it starts to decrease rapidly before a small continuous lift due to the repulsion inhabited in the SQM modeling. Then $\sim12\rho_0$, when it lowers to $\sim0.5$, $\Gamma$ increases due to the change from two phases to a single phase, and approaches the pure quark matter result (some value lower than 1) $\sim16\rho_0$. The increase of hadron-quark interface tension generally lowers the first peak and enhances the second peak simultaneously. Finally, for large $\sigma$ like $50~\rm MeV/fm^2$, only the second peak is present, similar to the Maxwell case and the pure quark matter case. The detailed variations for mixed-phase under various conditions depend mainly on the competition between the softening due to the coexistence of two phases and the stiffening due to the pressure increase.

In the systematic study of \cite{2020NatPh..16..907A}, an evident decrease of the adiabatic index is regarded as a signature of hadron-quark phase transition. However, the $\Gamma$ decrease can be achieved with only one phase of SQM, by using one of our effective model of quark matter, the perturbation model (with in-medium bags), as one may notice in Fig.~\ref{fig:gamma}.
The masquerading of quark matter as hadron-quark mixed phase can also be seen in the $c_s$ study, for example, in the lower panel of Fig.~\ref{fig:hybrid}.
The behavior of the sound speed of hadron-quark mixed-phase resembles that of the pure SQM case in the intermediate density region of $\sim3\text{-}8\rho_0$.
A recent Bayesian analysis on HSs adopting the GW170817 and {\it NICER} PSR J0030+0451 data found a similar $c_s$ peak value of $\sim0.81c$~\cite{Li2021a} as in the previous section for quark matter.
As a consequence, the distinguishing between different states of dense matter (including the onset of phase transition) can hardly be achieved by the variations in the sound speed or the adiabatic index, according to the present study.

\subsection{New series of stiff QS EOSs in the perturbation model with in-medium bags}
\label{sec:mr}

In Fig.~\ref{fig:mr}, we show QSs's mass as a function of the radius (in the left panel) and as a function of the central density (in the right panel).
The calculations are done with all four effective quark matter models collected in the present work, i.e., the quasiparticle model~$(C_1,B_0)=(3.5,~50)$, equivparticle model~$(C,\sqrt{D})=(0.7, 129)$, and MIT$\alpha_\mathrm{s}^2$~$(B_{\rm eff},a_4)=(138,~0.6)$.
In particular, we apply here for the first time the pertrubation model to the self-bound QSs, for two representative cases: $(C_1,~B_0,~\Delta \mu)=(3.5, 40, 800)$ as well as (3.5, 40, $\infty$).

It is seen that the radii of most massive QSs lie between $\sim10~\rm km$ and $\sim12~\rm km$, with one exception in the equivparticle model due to very low surface density $\sim 0.1~\rm fm^{-3}$ mentioned before. 
In the equivparticle model, it is necessary to have a large radius to ensure a large maximum mass above the two-solar mass. Such high radius ($\sim 14~\rm km$) may have been excluded by the LIGO/Virgo observation of NS binary merger GW170817 if one supposes it originates from binary QS merger. 
The repulsive contribution from the in-medium bag in the perturbation model demonstrates a new way to achieve a large maximum mass with a small radius; for example, the mass is lifted from $1.8\Msun$ (when $\Delta\mu=\infty$) to $2.2\Msun$ (when $\Delta\mu=800~\rm MeV$) with a similar radius. 
One more merit of the new perturbation model (with in-medium bag) is achieving both a large maximum mass and a large surface density.
A too low QS surface density (below $\rho_0$) is not welcomed since in such a density realm, they should be confined inside hadrons. 
Further measurements of a small radius (especially for small pulsars) together with a large maximum mass would help justify this QS EOS model~\cite{2020JHEAp..28...19L} and the effective scaling of the bag parameter used in the model.

\section{Summary}

The sound speed $c_s=\sqrt{dP/d\rho}$ is a fundamental quantity for describing the matter state, and the causality limit has been used to set important bounds on dense matter EOS and NSs' maximum mass~\cite{1976Natur.259..377B,1974PhRvL..32..324R}. For example, the polytropic form of $P=(\rho-\rho_0)c^2+P_m$ matched smoothly to a realistic nuclear matter EOS \cite{1973NuPhA.207..298N} at nuclear saturation density $\rho_0$ ($P_m$ is a constant determined from the matching) gives an upper limit of $\sim4.8\Msun$ for the TOV mass. In this study, we explore the possibility of using the microphysical quantities (like $c_s$) to shed light on particle degree of freedom in cold, dense matter in the density region where no first-principle method can be presently applied.

We make use of various many-body frameworks for the modeling of pure nuclear matter and quark matter. Those models employed cover approximately the full range
of NS/QS EOS models regarding their stellar properties. One representative quark matter model, the perturbation model, is used for the study of hadron-quark deconfinement phase transition, together with two representative EOS models (TW99 and DDME2) for nuclear matter.

We mainly find a dissimilarity of the adiabatic index for pure nuclear matter and quark matter. And a large sound velocity  (i.e., a particular shape) may be necessary for dense matter to fulfill the two-solar-mass constraint: $\ge 0.68 c$ for QSs and $\ge 0.8 c$ for HSs. 
Correspondingly, both $\Gamma$ and $c_s$ can not effectively signify the matter's composition in intermediate densities relevant to compact stars. 
The complication also arises from additional nonperturbative effects included in the model calculation, which brings extra repulsion above $\sim5\rho_0$ and affects the predicted structures of QSs. As a result, a more compact QS is possible with a TOV mass as high as $\sim2.2\Msun$. 
It is a new series of QS EOSs that could bring interesting observational possibilities to study the EOS of dense QCD matter and the nonperturbative properties of QCD. Along this line, some studies have pointed out the difference in the dynamical stability between one phase stars and multi-phase stars~\cite{2018ApJ...860...12P}. Further efforts regarding the dynamical properties (like NS cooling~\cite{2020JHEAp..28...19L,2020PhRvC.101f5801S}, NS binary merger simulation, etc) of such compact objects, may be necessary for identifying the state of QCD matter at intermediate densities.

\begin{acknowledgements}
This work was supported by National SKA Program of China (No. 2020SKA0120300), the National Natural Science Foundation of China (Grant No. 11873040), the Youth Innovation Fund of Xiamen (Grant No. 3502Z20206061), Ningbo Natural Science Foundation (Grant No. 2019A610066), CAS ``Light of West China" Program No.~2018-XBQNXZ-B-025 and Tianshan Youth Program (No. 2018Q039). The support provided by China Scholarship Council during a visit of C.-J. X. to JAEA is acknowledged. The computation for this work was supported in part by the HPC Cluster of SKLTP/ITP-CAS and the Supercomputing Center, CNIC, of the CAS.
\end{acknowledgements}


\end{document}